%% file: main.tex
\documentclass[10pt,conference]{IEEEtran}
\usepackage{graphicx}

\IEEEoverridecommandlockouts
\input{packages}

\begin{document}

\title{Finding Anomalies in Scratch Programming Assignments}
\title{Anomalies in Scratch Programming Assignments}
\title{Anomaly Detection in Scratch Programming Assignments}
\title{Finding Anomalies in Scratch Assignments}
\title{Anomaly Detection in Scratch Assignments}

\author{\IEEEauthorblockN{Nina Körber}
\IEEEauthorblockA{
\textit{University of Passau}\\
Passau, Germany}
}

\maketitle

\input{content/abstract}


\input{content/introduction-motivation}
\input{content/background}
\input{content/setup-evaluation}

\balance

\bibliographystyle{IEEEtranS}
\bibliography{references}
 
\end{document}

%% file: packages.tex
\usepackage{microtype}
\usepackage{graphicx}
\usepackage{booktabs}
\usepackage{siunitx}
\usepackage{paralist}
\usepackage{multirow}
\usepackage{array}
\usepackage{balance}
\usepackage{fontawesome}
\usepackage[para]{footmisc}
\usepackage{cite}
\usepackage{makecell}
\usepackage{hyperref}
\usepackage[hyphenbreaks]{breakurl}

\usepackage{xcolor}

\usepackage{wrapfig}
\usepackage[caption=false,font=footnotesize]{subfig}

\usepackage[T1]{fontenc}
\usepackage{float}
\usepackage[utf8]{inputenc}
\usepackage{subfiles}
\usepackage{caption}
\usepackage[noend]{algorithmic}
\usepackage{algorithm}

\usepackage{tocbasic}
\usepackage{multicol}
\usepackage[]{scrlayer-scrpage}

\usepackage{syntax}
\renewcommand{\litleft}{\sf \begin{tikz}[baseline=(X.base)]\node [draw=gray!40,fill=gray!0,semithick,rectangle,inner sep=1pt, minimum size=1em, outer sep=0pt, rounded corners=2pt] (X) }
\renewcommand{\litright}{;\end{tikz} \normalfont}

\usepackage{enumitem}
\usepackage{xspace}
\usepackage{xargs}
\usepackage{amsthm}
\usepackage{amsmath}
\usepackage{amsfonts}
\usepackage{amssymb}
\usepackage{pifont}
\newcommand\definetool[2]{\newcommand{#1}{{\textsc{#2}}\xspace}}
\definetool{\scratch}{Scratch}
\definetool{\whisker}{Whisker}
\definetool{\litterbox}{LitterBox}
\definetool{\bastet}{Bastet}
\definetool{\hairball}{Hairball}
\definetool{\drscratch}{Dr. Scratch}
\definetool{\qualityhound}{Quality Hound}
\definetool{\jadet}{Jadet}
\definetool{\aumextr}{GeSMo}
\definetool{\oumextractor}{OUMExtractor}
\definetool{\aums}{AUMs}
\definetool{\aum}{AUM}
\definetool{\oums}{OUMs}
\definetool{\oum}{OUM}
\definetool{\java}{Java}

\newcommand{\sub}{\textsubscript}

\usepackage{multirow}
\usepackage{array}
\newlength{\defaulttabcolsep}
\definecolor{OliveGreen}{rgb}{0,0.6,0}
\usepackage[colorinlistoftodos,prependcaption,textsize=small]{todonotes}
\newcommandx{\missing}[2][1=]{\todo[linecolor=red,backgroundcolor=red!25,bordercolor=red,#1]{#2}}
\newcommandx{\unsure}[2][1=]{\todo[linecolor=orange,backgroundcolor=orange!25,bordercolor=orange,#1]{#2}}
\newcommandx{\change}[2][1=]{\todo[linecolor=blue,backgroundcolor=blue!25,bordercolor=blue,#1]{#2}}
\newcommandx{\info}[2][1=]{\todo[linecolor=OliveGreen,backgroundcolor=OliveGreen!25,bordercolor=OliveGreen,#1]{#2}}

\usepackage{fbox}
\usepackage[formats]{listings}
\definecolor{lightgray}{rgb}{.9,.9,.9}
\definecolor{darkgray}{rgb}{.4,.4,.4}
\definecolor{purple}{rgb}{0.65, 0.12, 0.82}
\lstdefinelanguage{JavaScript}{
  keywords={typeof, new, true, false, catch, function, return, null, catch, switch, var, if, in, while, do, else, case, break},
  keywordstyle=\color{blue}\bfseries,
  ndkeywords={class, export, boolean, throw, implements, import, this},
  ndkeywordstyle=\color{darkgray}\bfseries,
  identifierstyle=\color{black},
  sensitive=false,
  comment=[l]{//},
  morecomment=[s]{/*}{*/},
  commentstyle=\color{purple}\ttfamily,
  stringstyle=\color{red}\ttfamily,
  morestring=[b]',
  morestring=[b]"
}

\usepackage{tikz}
\usepackage{pgflibraryshapes}
\usetikzlibrary{arrows}
\usetikzlibrary{chains}
\usetikzlibrary{fadings}
\usetikzlibrary{matrix}
\usetikzlibrary{fit}
\usetikzlibrary{positioning}
\usetikzlibrary{shapes}
\usetikzlibrary{automata}

\tikzset{
 ctrlstate/.style = {state,align=center,inner sep=2pt, minimum size=2mm},
 cfastate/.style = {ctrlstate},
 cfatargetstate/.style = {cfastate, double},
 compstate/.style = {cfastate, rounded rectangle, minimum height=5mm, minimum width=3em, inner sep=3pt},
 concept/.style = {cfastate, inner sep=3pt, fill=gray!50, rectangle, 
        minimum width=17mm, minimum height=9mm, draw=gray!6 },
 crosscutting/.style = {concept, rounded rectangle, fill=gray!30},
 conceptstate/.style = {concept, rounded rectangle, fill=gray!30, draw=gray!90, minimum width=20mm},
 inputstate/.style = {concept, rectangle, fill=gray!10, draw=gray!90, minimum width=20mm, inner sep=3pt},
 explain/.style = {circle, draw=gray!30, line width=1mm, minimum size=7mm},
 one/.style = {fill=blue!70,draw=blue!70},
 two/.style = {fill=red!50,draw=red!50},
 abststate/.style = {rectangle,align=center,inner sep=2pt,minimum size=3.5mm,fill=gray!0, draw=gray!90},
 line/.style = {draw},
 trans/.style = {draw,semithick,->,shorten >=1pt,>=stealth'},
 missing/.style = {draw=red,densely dotted,fill=red,semithick,->,shorten >=1pt,>=stealth'},
 ctrans/.style = {draw,very thick,->,shorten >=1pt,>=stealth',draw=gray!90},
 epsilon/.style = {trans,dashed},
 strengthen/.style = {draw=gray!30,semithick,double,shorten >=1pt,>=stealth',line width=1mm},
}

\newcommand{\blockleft}{\begin{mbox}\sf\begin{tikz}[baseline=(X.base)]\node[draw=black!60,fill=black!3,semithick,rectangle,inner sep=1pt, minimum size=1em, outer sep=0pt, rounded corners=1pt] (X)}%
\newcommand{\blockright}{;\end{tikz}\normalfont\end{mbox}}%

\newcommand{\mblockleft}{\begin{mbox}\sf\begin{tikz}[baseline=(X.base)]\node[draw=red!60,densely dotted,fill=red!3,semithick,rectangle,inner sep=1pt, minimum size=1em, outer sep=0pt, rounded corners=1pt] (X)}%
\newcommand{\mblockright}{;\end{tikz}\normalfont\end{mbox}}%

%% file: content/abstract.tex
\begin{abstract}
%
%
For teachers, automated tool support for debugging and assessing
their students' programming assignments is a great help in their everyday business.
%
%
For block-based programming languages which are commonly used to introduce younger
learners to programming, testing frameworks and other software analysis tools exist, but
require manual work such as writing test suites or formal specifications. However,
most of the teachers using languages like \scratch are not trained
for or experienced in this kind of task. Linters do not require manual work but are limited
to generic bugs and therefore miss potential task-specific bugs in student solutions.
%
%
In prior work, we proposed the use of anomaly detection to find
project-specific bugs in sets of student programming assignments automatically, without
any additional manual labour required from the teachers' side.
Evaluation on student solutions for typical programming assignments showed
that anomaly detection is a reliable way to locate bugs in a data set of student
programs. In this paper, we enhance our initial approach by lowering the abstraction level.
The results suggest that the lower abstraction level can focus anomaly
detection on the relevant parts of the programs.
\end{abstract}

%% file: content/introduction-motivation.tex
\section{Introduction and Motivation}\label{sec:introduction}

Teachers often assess the progress of their students'
programming and computational thinking skills with
tasks the students have to solve practically by coding. In this process, one of the tasks of the teacher
is to assess and evaluate the programs produced to provide feedback or grade the
students. In practice, this task is challenging, as there are many ways to solve a
programming assignment and an unlimited variety of potential bugs, making it especially hard if a teacher is inexperienced~\cite{Sentance.2017}.
For text-based programming languages, there are many ways in which teachers can
utilize software analysis tools to help with this process, for example by running tests,
using a linter or writing specifications to check against.

However, there is an increasing demand to teach programming to elementary
students, and the typical way to introduce young learners is to use block-based
programming languages such as \scratch~\cite{maloney2010}. For these languages, there rarely exist
tools that help teachers with assessing their students' code. The few tools that
exist are either limited because they statically check for a predefined set of
bug patterns~\cite{bugpatterns} or require the elementary school teachers to be trained in writing
formal specifications~\cite{VerifiedFromScratch} or test suites~\cite{whisker}, which can be hard even for experienced programmers.

To address this problem, \emph{anomaly detection} can
help teachers assess and evaluate student assignments~\cite{koerber2021}.
Anomaly detection is a static software analysis technique that has been
successful at finding bugs in large projects using text-based programming languages~\cite{jadet}.
The idea is to learn common coding patterns, rare deviations of which, \textit{anomalies}, potentially
hint at bugs in the code. Regular student solutions at elementary schools
usually do not have a sufficient size to learn patterns from a single solution.
However, there usually is one task for a whole class, resulting in a set of
student solutions that consist of high-level code constructs. We use this set of
student solutions to learn common student coding patterns and to find anomalies.

In prior work, we adapted and extended an approach to anomaly detection from the context of object-oriented
programming based on formal concept analysis~\cite{jadet} to the context of \scratch.
We mined and evaluated anomalies of six data sets
consisting of different \scratch solutions created by students~\cite{koerber2021}. To further explore
the potential of anomaly detection in \scratch, in this paper we propose
\textit{actor specific} anomaly detection, which groups scripts by the actor
(sprite or background) they belong to. Our results show that both the original and the actor specific
approach are reliable ways to find generic as well as project-specific bugs in student code,
without any manual work from teachers. We also found that the actor specific approach
can focus anomaly detection on the relevant parts of the student solutions and its
effects are worth further exploration.
%

%% file: content/background.tex
\section{Anomaly Detection in Scratch: Approach}
\label{ch:background}
Several approaches for anomaly detection for text-based programming languages have been proposed in the
literature~\cite{dynamine,li2005pr,eisenbarth,grouminer,wasylkowski2011mining,jadet}.
The common idea is that correct code is the rule rather than the exception---common
behaviour likely is correct behaviour and rare deviations of it likely are wrong.
Our approach is based on the \java anomaly detector \jadet~\cite{jadet}, which mines
coding patterns related to \java objects. Therefore, the usage of every \java object
is analyzed. For every \java class, common patterns of method calls on instances
of this class are mined. If there are rare deviations of these patterns, i.e., an
object is not used the way objects of its class usually are, the unusual behaviour is marked
and reported. Pattern mining relies on graph models of \java objects, but the actual
algorithm used for mining patterns and their anomalies from these models is independent of
the programming language underlying the graph models. This allows us to reuse
these algorithms without major adaptations. The main difference of our approach
is that we mine patterns and violations from graph models of \scratch scripts.
We implemented the extraction of these in the \litterbox analysis framework for \scratch~\cite{bugpatterns}.

\begin{figure*}[t]
  \includegraphics[width=\textwidth]{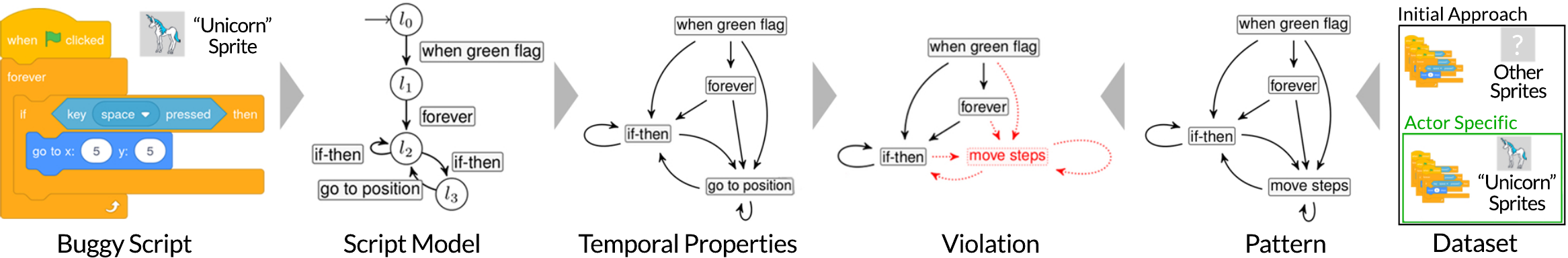}
  \caption{\label{fig:overview}Anomaly detection in \scratch is able to find project-specific bugs. In this example the \textit{move steps} block is missing and instead a \textit{go to position} block is used in the script on the left in order to move the sprite when the space key is pressed.}
\vspace{-0.5em}
\end{figure*}
\newcommand{\nats}{\ensuremath{\mathbb{N}}\xspace}%

Figure~\ref{fig:overview} presents the four essential phases of anomaly detection in \scratch~\cite{koerber2021}.
\subsubsection{Script Model Generation} Every script in the data set is converted
into a \textit{script model} where nodes are control locations in the script that can
be reached by executing the blocks on the transitions, starting from the initial
location l\sub{0}.
\subsubsection{Temporal Properties} Every script model is translated into its \textit{temporal
properties}, i.e., pairs of blocks that occur one after the other in the control flow
of a script.
\subsubsection{Pattern Mining}\textit{Patterns} are sets of temporal properties that occur in at
least \textit{k} scripts, where \textit{k}~$\in~\nats$ is a minimum support threshold.
Patterns are mined from the temporal properties of all scripts in the student solutions for the task via frequent itemset mining.
\subsubsection{Anomaly Mining} A script \textit{violates} a pattern if the pattern
is not a subset of the temporal properties of the script. A violation will only
be reported as \textit{anomaly} if there are many scripts that adhere to, \textit{support}, the violated pattern compared
to scripts that violate the pattern in the exact same way the script at hand does, i.e., the \textit{confidence}
of the violation is high.
%

%% file: content/setup-evaluation.tex
\section{Initial Results}\label{sec:eval}
We evaluated our approach on six data sets of student solutions for \scratch
assignments (799 solutions in total), one of them being an open task where students were free to implement
what they wanted to~\cite{koerber2021}. All of these solutions were created from students
aged 8-13 during programming sessions with qualified teachers.
We found that anomalies can be found,
and that buggy solutions more likely have more anomalies reported. We classified
the top ten anomalies for every data set into the categories
\textit{defective},
\textit{smelly} and \textit{non-defective}.
In total, we classified 60 anomalies, 46 of which hinted at defective code,
4 at smelly code and 10 at non-defective work.
While this is an encouraging result, teachers might benefit if we could better
separate non-defective from defective anomalies.
\newcommand{\toollink}{\smwouldall{\texttt{\url{https://github.com/se2p/scratch-anomalies}}}}%

\section{Actor Specific Anomaly Detection}
\subsection{Motivation and Differences to the Initial Approach}
One of the reasons for the success of \jadet is the distinction
between the classes objects belong to; anomalies are \textit{class specific}~\cite{jadet}.
Our initial approach for anomaly detection in \scratch abstracts away from the actor (sprite or background)
of a script; the anomalies are \textit{actor agnostic} (AA). However, actors in \scratch are the block-based equivalent to
classes in \java, so it might be beneficial to keep the information about actors.
To explore these potential benefits, we implemented an anomaly detection variant we call
\textit{actor specific} (AS) anomaly detection: We group scripts by the purpose of their actors for pattern and anomaly mining.
For open tasks, grouping the actors by purpose is a non-trivial task that requires
matching the scripts inside the actors. For tasks with tighter constraints, we assume that matching actors by their
names approximates the similarity of their purposes sufficiently.

\subsection{Pilot Study}
We mined AS anomalies on one of the tightly specified tasks of our original data set,
the \textit{cat task}.
The AA approach found 91 violations and reported 30 of them in our
initial experiments~\cite{koerber2021}. Using the same
parameters, the AS approach found 54 violations and reported 20 of them,
as the others were below the confidence threshold configured.

For all reported anomalies including the ones reported in both approaches, the
confidence values changed.
The confidence value of a violation depends on the number of scripts that
support a pattern or violate it in the same way the violation at hand does.
Consequently, the exclusion of scripts that belong to other actors changes the
confidence values and therefore the ranking of the anomalies.
The lower number of AS violations found indicates that the approach successfully avoided
non-defective anomalies. To investigate whether this is actually the case, as part of our experiments we
manually inspected the 30+20 anomalies. Out of these, 12 were identical in both sets.

Of the 30 AA anomalies, 8 hint at scripts that are located in irrelevant actors
that were supposed to be left empty or that were added by the students.
In general, we noticed that these anomalies rarely influence the correctness of
a solution; it mostly depends on the implementation
quality of the relevant actors, i.e., those that are supposed to be programmed.
While these anomalies may be of interest because they hint at creative solutions,
they do not help debugging problems.
As a clear improvement for the application in debugging, the AS approach
yields only anomalies that hint at code in relevant actors. Furthermore, this effect
serves as an explanation for the reduced number of AS anomalies found.
\section{Summary}
Anomaly detection in \scratch is a useful approach to help teachers
with assessing their students' code. This research field is still young and
a first step to improve the initial approach by mining AS anomalies
yielded promising results that are worth investigating in further experiments.
